\documentclass[aps,twocolumn,prl,superscriptaddress,
preprintnumbers,floatfix,tightenlines, 10pt]{revtex4-2} 
\usepackage{amsmath,amsfonts,amssymb,amscd,amsxtra,amsthm}
\usepackage{dsfont}
\usepackage{graphicx}  % Include figure files
\usepackage{epstopdf}
\usepackage{dcolumn}  % Align table columns on decimal point
\usepackage{bm}          % bold math
\usepackage{slashed}
\usepackage[utf8]{inputenc}
\usepackage{booktabs}
\usepackage[normalem]{ulem} % \sout{old text} for strikeout
\usepackage[dvipsnames]{xcolor} % For blue in-text comments and
\usepackage{enumitem}
\usepackage{array}
\usepackage{slashed}
\usepackage{tikz}
\usepackage{float}
\usepackage{multirow}
\usepackage{booktabs} 
\usepackage{slashed}

\renewcommand\sout{\bgroup \color{red} \ULdepth=-.5ex \ULset}

\begin{document}
\preprint{INHA-NTG-05/2022}
\title{Triple Regge exchange and transverse single-spin asymmetries 
of the very forward neutral pion production in polarized $p+p$
collisions} 
%------------------------------------------------
\author{Hee-Jin Kim}
\affiliation{Department of Physics, Inha University, Incheon 22212,
Republic of Korea}
\author{Samson Clymton}
\affiliation{Department of Physics, Inha University, Incheon 22212,
Republic of Korea}
\author{Hyun-Chul Kim}
\affiliation{Department of Physics, Inha University, Incheon 22212,
Republic of Korea}
\affiliation{School of Physics, Korea Institute for Advanced Study
(KIAS), Seoul 02455, Republic of Korea}
\date{\today}
\begin{abstract}
Recently, the RHICf Collaboration measured the transverse single-spin
asymmetries of the very forward neutral pion in polarized $p+p$
collisions at $\sqrt{s}=510$ GeV, produced at large pseudorapidity 
($\eta\gtrsim 6$). The data show large asymmetries both in
longitudinal momentum fraction $x_F$ and transverse momentum $p_T$ at
$p_T<1\,\mathrm{GeV}/c$. Employing baryonic triple Regge exchanges, we
describe the complete RHICf data for the first time and show that the
neutral pion production at low $p_T$ can be interpreted as a
diffractive one. 
\end{abstract}
\maketitle

% ======================================
\textit{Introduction} -- The spin of the nucleon has been one of the
 most crucial issues in hadronic physics since the ``spin crisis''
 caused by the EMC experiment~\cite{EuropeanMuon:1987isl}. Since the
 nucleon consists of not only three valence quarks but also other
 partons such as antiquarks and gluons, the nucleon spin should
 originate from the partons inside it and their orbital angular
 momenta~\cite{Ji:1996ek}. ``\emph{How does the spin of the nucleon
 arise?}'' This profound question motivated the future plan for
 the Electron-Ion Collider (EIC)~\cite{EIC}. 
Meanwhile,  the transverse spin of the nucleon provides yet
 another aspect to  the internal structure of the nucleon. The
 transverse  momentum-dependent functions(TMDs) together with the
 generalized  parton distributions(GPDs) furnish the multi-faceted
 aspect of the  structure of the polarized nucleon in
 the transverse plane~(see recent reviews~
 \cite{Accardi:2012qut,Anselmino:2020vlp}). Furthermore, sizable
 transverse single-spin asymmetries (TSSA) of the neutral pion in
 inclusive $pp$ collisions have been continuously reported well over
 decades~\cite{Klem:1976ui, E581:1991eys, STAR:2008ixi,
 PHENIX:2013wle, PHENIX:2020mft, STAR:2020nnl}(see also recent
 reviews~\cite{Aschenauer:2016our, Anselmino:2020vlp}). Since
 the experimental data from the PHENIX and STAR Collaborations were
 obtained at higher values of the transverse momentum ($p_T\gtrsim 2
 \mathrm{GeV}/c)$ in the mid-rapidity coverage, where the
 pseudorapidity is given as $2<\eta<4$~\cite{STAR:2008ixi, 
 PHENIX:2013wle, PHENIX:2020mft, STAR:2020nnl}, QCD-based approaches
 have been employed such as the TMDs~
 \cite{Sivers:1989cc,Collins:1992kk,Collins:1993kq} and
 collinear twist-3 factorization~\cite{Efremov:1981sh, Eguchi:2006mc,
 Qiu:1991wg, Qiu:1991pp, Gamberg:2013kla, Kanazawa:2014dca,
 Gamberg:2017gle} to describe the experimental data. The Jefferson
 Lab Angular Momentum(JAM) Collaboration~\cite{Cammarota:2020qcw} has
 carried out the simultaneous QCD global analysis, considering the
 data on the TSSA from various high-energy processes. 

The TSSA at low transverse momentum in the large pseudorapidity
display the nonperturbative diffractive 
nature. The RHICf Collaboration measured the TSSA of the neutral pion
in transversely polarized $p^\uparrow+p$ collision at $\sqrt{s}=510$
GeV and reported that the TSSA increased rapidly as functions of both
the longitudinal momentum fraction $x_F$ and low transverse momentum
$p_T$ ($p_T<1\,\mathrm{GeV}/c$) at the pseudorapidity larger than 6
($\eta>6$)~\cite{RHIC-f:2020koe}. The RHICf experiment data posed a
question of whether the large values of the TSSA of $\pi^0$ are due to
diffractive scattering: The values of TSSA rise as $p_T$ increases and
reach around 25~\% at $p_T\simeq 0.8\,\mathrm{GeV}/c$. The dependence
of the longitudinal momentum fraction or the Feynman-$x$ variable
($x_F$) reveals even a drastic feature. In this Letter, we will answer
for the first time the question addressed by the RHICf Collaboration:
Considering the $p^\uparrow+p\to \pi^0+X$ process at low $p_T$ as
diffractive scattering and introducing the baryonic triple-Regge
exchanges, we explain the RHICf data very well.    

\vspace{1em}
\emph{Triple-Regge exchange} --
The applications of the Regge approach to inclusive
hadronic reactions is dated back to the 1970s~\cite{Mueller:1970fa,
  Mueller:1971ez, Abarbanel:1971za, Phillips:1972ys, Field:1974fg,
  Chu:1975gv, Paige:1976iy}. Mueller generalized the optical
theorem for inclusive reactions: the differential cross section for
the two-body inclusive reaction, $a+b\to c+X$, can be written in terms
of discontinuity of the three-body process $ab\bar{c}\to ab\bar{c}$
along the missing mass $M_X^2=(p_a+p_b-p_c)^2$. It was shown that the 
three-body amplitude has the Regge singularities similar to those for
the two-body process~\cite{Mueller:1970fa, Collins:1977}. The
triple-Regge exchange is obtained from an asymptotic behavior of the
Mueller amplitude in the kinematic boundary. It was shown that the
unpolarized cross section was successfully described by triple-Regge
pole contributions~\cite{Field:1974fg}. The triple-Regge formalism is 
still employed as a robust tool for understanding diffractive
processes~\cite{Fiore:1999zb,Godizov:2015hbm,Levin:2018qxa}. 

\begin{figure*}[htp]
\centering
\includegraphics[scale=.8]{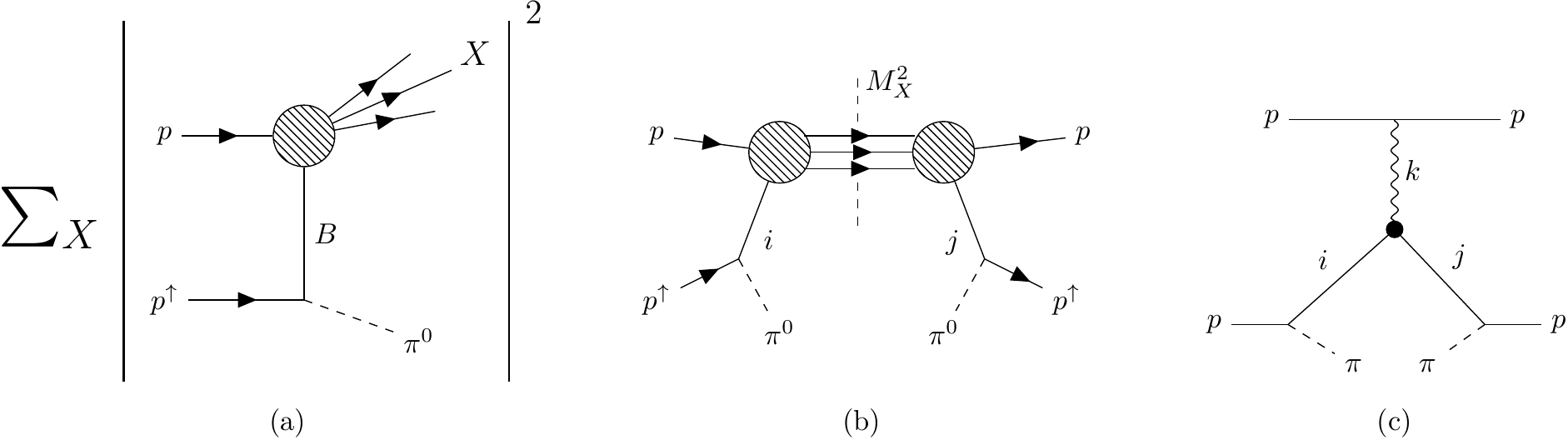}
\caption{Diagrammatric representation of $d\sigma^h$ and
triple-Regge diagram. $d\sigma^h$ is proportional to diagram (a).  
It can be approximated to (b) in the high energy region. Diagram (c) 
illustrates that with triple-Regge exchange. When $M_X^2$ is 
large, (b) can be replaced by (c).}
\label{fig:1}
\end{figure*}

On the other hand, it was anticipated that the nondiagonal
triple-Regge pole diagram would be able to present the polarization 
effects~\cite{Russ:1975qt,Paige:1976iy}. However, the statistics of
the experiments was very poor, so that no significant data were 
reported at that time. Only very recently, the RHICf experiment
accomplished a measurement of the TSSA in the very forward
direction~\cite{RHIC-f:2020koe}. Since the final  
particles have very high pseudorapidities and low transverse
momenta, one can use Regge-exchange of the initial proton as shown in
Fig.~\ref{fig:1} (a). As mentioned previously, the generalized optical
theorem leads to Fig.~\ref{fig:1} (b) with the discontinuity on the
complex $M_X^2$ plane. When $M_X^2$ is sufficiently large, $ip\to
jp$ scattering can be also expressed as a Regge pole. Thus one can
consider triple-Regge exchange to derive the TSSA as drawn in
Fig.~\ref{fig:1} (c). 
We extend the formalism in Ref.~\cite{Paige:1976iy} with baryon
Regge trjectories introduced. Note that the spin should be transfered 
by the baryon reggeons, since the produced pion does not carry any
spin from the polarized proton. 

The Lorentz-invariant differential cross section for the inclusive 
reaction $p+p^h\to \pi^0+X$ in the high energy limit is given by
\begin{align} \label{eq:LIDC}
d\sigma^h \equiv E\frac{d^3\sigma^h}{d^3\bm{p}} =
\frac 1{s} \sum \left|A_{p\to \pi^0}^
\text{tot}(s,p_T;h)\right|^2,
\end{align}
where $h$ is the helicity direction of the polarized proton
beam. $s$ denotes the square of the energy in the center of mass (CM) 
framework, which is one of the Mandelstam variables. $p_T$ stands for
the transverse momentum. 
When the energy is high enough to apply the Regge formalism,
we can use the generealized optical theorem by
Mueller~\cite{Mueller:1970fa} to express
$d\sigma^h$ in terms of two reggeon exchange $i$ and $j$ and the
scattering of the unpolarized proton and the reggon with the energy
$M_X^2$ as depicted in Fig.~\ref{fig:1}. 

In the limit $M_X^2\to \infty$, the discontinuity of the
$i+p\to j+p$ scattering will follow the Regge behavior as
\begin{align} \label{}
\mathrm{Disc}\, A_{ip\to jp}(M_X^2) = \sum_{k} G^
{ij}_{k}(t) \gamma^{pp}_k(0)
\left(\frac{M_X^2}{s_0}\right)^{\alpha_k(0)},
\end{align}
where $G_k^{ij}(t)$ represents the triple-reggeon coupling given as a
function of $t$ that is a square of the momentum transfer (one of the
Mandeltam variables), corresponding to the black blob in
Fig.~\ref{fig:1}(c), which will be discussed later. $\gamma_{k}^{pp}$
is the vertex function for the $ppk$ vertex in
Fig.~\ref{fig:1}(c). $s_0$ denotes the scale parameter, which is
traditionally given to be around 1 $\mathrm{GeV}^2$. 
Then $d\sigma^h$ is written as
\begin{align} \label{dsigmah_TR}
d\sigma^h \sim 
\beta_{h\lambda}^i \beta_{h\mu}^{j*} 
\mathcal{P}_i \mathcal{P}_j^* G^{ij}_k(t) \gamma^{pp}_k(0)
\left(\frac{M_X^2}{s_0}\right)^{\alpha_k(0)},
\end{align}
where $\beta^{i(j)}$ stands for the residue of $i$($j$) exchange and
$\mathcal{P}_{i}(t)$ is the reggeon propagator defined
as~\cite{Collins:1977,Donnachie:2002} 
\begin{align} \label{}
\mathcal{P}_i(t) \equiv \alpha_B'
\xi_i^\pm(t) \Gamma\left(J_i - \alpha_i(t)\right)
\left(1-x_F\right)^{-\alpha_i(t)}.
\end{align}
Here $\alpha_i(t)$ and $J_i$ denote respectively the Regge trajectory
and the spin for exchange particle $i$. The signature factor
is given as 
\begin{align} \label{}
\xi_i(t) = \frac{1+\tau_i\exp\{-i\pi(\alpha_i(t)-0.5)\}}{2},
\end{align}
where $\tau_i$ represents the signature of the corresponding reggeon,
i.e., $\tau_i = (-1)^{J_i-1/2}$.

We introduce the proton, the $\Delta(1232)$ isobar, the excited baryon
$N^*(1520)$, and the $\Delta(1600)$ isobar with negative parity to
derive the TSSA of very forward pion production as shown in
Fig~\ref{fig:1}. Since the Regge approach does not provide the vertex
structure, we need to employ the effective Lagrangians for $NN\pi$,
$N\Delta\pi$, $NN^*\pi$ and $N\Delta^*\pi$ given as
\begin{align} 
\mathcal{L}_{\pi NN} &= -\frac{f_{\pi NN}}{m_\pi} \bar{\psi}
\gamma_\mu \gamma_5
\bm{\tau} \cdot \psi \partial^\mu \bm{\pi}, \cr
\mathcal{L}_{\pi NN^*} &= -i\frac{f_{\pi NN^*}}{m_\pi}
\bar{\psi}_{N^*}^\mu (g_{\mu\nu} + a\gamma_\mu \gamma_\nu)
\gamma_5 \bm{T} \cdot \psi \partial^\nu \bm{\pi}, \cr
\mathcal{L}_{pi N\Delta} &= -\frac{f_{\pi N\Delta}}{m_\pi}
\bar{\psi}_
\Delta^\mu (g_{\mu\nu} + a\gamma_\mu \gamma_\nu) \bm{T}
\cdot \psi\partial^\nu \bm{\pi}  , \cr
\mathcal{L}_{\pi N\Delta^*} &= -\frac{f_{\pi N\Delta^*}}
{m_\pi} \bar{\psi}_
{\Delta^*}^\mu (g_{\mu\nu} + a\gamma_\mu \gamma_\nu)
\bm{T} \cdot  \psi\partial^\nu \bm{\pi},
\label{eq:6}
\end{align}
where $\psi$, $\psi^\mu$ and $\pi$ denote  respectively the Dirac,
Rarita-Schwinger, and pseudoscalar fields for the nucleon, $\Delta$
isobar, and the pion. $f_{\pi NN}$, $f_{\pi NN^*}$, $f_{\pi N\Delta}$,
and $f_{\pi N\Delta^*}$ designate the strong coupling constants for
the corresponding vertices and $m_\pi$ is the pion mass. $\bm{\tau}$
represents the Pauli matrix for the spin 1/2 isospin
operator and $\bm{T}$ stands for the isospin transition operator from
isospin 1/2 to 3/2 states. $g_{\mu\nu}$ is the metric tensor
$g_{\mu\nu}=\mathrm{diag}(1,-1,-1,-1)$ and $a$ the off-shell
parameter for the spin 3/2 baryon. 

The Regge factorization implies that the Born amplitudes of
the one-particle exchange (OPE) can be subdivided into
residues for each vertex and reggeon propagator. The
proton-baryon-pion vertex functions are computed from the
given effective Lagrangian respectively as follows:
\begin{align} \label{eq:residue_functions}
\beta_{\lambda\lambda'}^N(p_T) &= \bar{u}_N(\lambda',q)
                \slashed{k} \gamma_5 u_p(\lambda,p), \cr
\beta_{\lambda\lambda'}^{N^*}(p_T) &= 
    i\bar{u}^\mu_{N^*}(\lambda',q)   (k_\mu + a\gamma_\mu \slashed{k})
   \gamma_5 u_p(\lambda,p)),           \cr 
\beta_{\lambda\lambda'}^\Delta(p_T) &= 
    \bar{u}^\mu_\Delta(\lambda',q) (k_\mu + a\gamma_\mu \slashed{k}) 
    u_p(\lambda, p), \cr   
\beta_{\lambda\lambda'}^{\Delta^*}(p_T) &= 
    \bar{u}^\mu_{\Delta^*}(\lambda',q)  
    (k_\mu + a\gamma_\mu \slashed{k}) u_p(\lambda,p),
\end{align}
where $p$, $k$, and $q$ are the four-momenta of the proton, pion and  
the exchaged reggeon. $\lambda$ and $\lambda'$ denote the helicities 
of the baryons. For simplicity, we will switch off the off-shell 
parameter $a$ in Eq.~\eqref{eq:6}. Note that
$\beta_{\lambda\lambda'}^i$ should be real-valued functions, so the
signature factor determines the phase of 
$d\sigma$. It plays a crucial role in deriving the TSSA. Note that by
the Abarbanel-Gross theorem the triple-Regge exchange does not yield 
the TSSA in the backward direction, which agrees with the RHICf
experimental data~\cite{RHIC-f:2020koe}. The TSSA in the backward
direction are almost consistent with zero~\cite{RHIC-f:2020koe}. 
\vspace{0.5em}

\emph{Transverse single spin asymmetry} ---\quad
The transverse single spin asmmetry is defined by the ratio of 
spin-dependent and spin-average differential cross section:
\begin{align} \label{eq:defaysm}
A_N = \frac{d\Delta\sigma_\perp}{d\sigma}
=\frac{d\sigma^\uparrow - d\sigma^\downarrow}
{d\sigma^\uparrow + d\sigma^\downarrow},
\end{align}
where $\uparrow(\downarrow)$ indicates the polarization of
the proton in the transverse direction. Inserting Eq.~\eqref{eq:LIDC} 
into Eq.~\eqref{eq:defaysm}, we can straightforwardly compute $A_N$. 
Before we derive the explicit expression for $A_N$, we
discuss the parity invariance of $\beta^i$ that will provide two
constraints. Firstly, $d\sigma$ vanishes if state $k$ has  
unnatural parity. A matrix element that consists of two fermions
(1, 2) and a spinless particle (3) obeys the following parity relation
$\beta_{\lambda_1 \lambda_2}^3 = \eta_1 \eta_2 \eta_3 (-)^ 
{\lambda_1 - \lambda_2} \beta_{-\lambda_1, -\lambda_2}^3$,
where $\eta_i$ denotes the naturality of a particle $i$. Since the 
proton has a natural parity, we have $\eta_1\eta_2=+1$. So, the
residue of the $ppk$ vertex satisfies the parity relation
$\beta_{\lambda \mu}^k = \eta_k
(-)^{\lambda-\mu}\beta_{-\lambda,-\mu}^k$. 
It leads to $\gamma_k^{pp}(0) = \sum_{\nu} \beta_{\nu\nu}^k = (1+\eta_k) 
\beta_{++}^k$, which becomes zero when state $k$ has
unnatural parity. For example, the following particles such as $\pi$,
$a_1$, and $\omega$, etc. have unnatural parity. Thus only the
particles with natural parity, $k=\mathds{P},\rho,a_2$,
etc., can contribute to $A_N$. Here $\mathds{P}$ represents the
pomeron. Following Eq.~\eqref{dsigmah_TR}, we find that 
pomeron exchange contributes to $d\sigma^h$ dominantly over other
meson exchanges that have $\alpha_k(0)$ less than $0.5$.

Secondly, $d\Delta\sigma_\perp$ vanishes when $i$ and $j$
have opposite naturality each other. As mentioned previously, 
the unpolarized proton does not take across any information on its
spin to the final state. This implies that $k$ exchange will not
affect the spin polarization of particle $i$. Since the
transversely polarized state is expressed in terms of
positive and negative helicity states quantized along the
$z$-axis: $|\uparrow\,\rangle = (|+\rangle + i |-\rangle)/
\sqrt{2}$ and $|\downarrow\,\rangle = (|+\rangle
-i|-\rangle)/\sqrt{2}$, the residue functions in
Eq.~\eqref{dsigmah_TR} are expressed as  
\begin{align} \label{eq:ytoz2}
\beta_{\uparrow \lambda}^i = \frac 1{\sqrt{2}}
(\beta_{+\lambda}^i +i \beta_{-\lambda}^i),
\;\;\;
\beta_{\downarrow \lambda}^i = \frac 1{\sqrt{2}}
(\beta_{+\lambda}^i -i \beta_{-\lambda}^i).
\end{align}
Using the fact that $\eta_p = +$ and $\eta_\pi = -$, we 
observe 
\begin{align}
d\Delta\sigma_\perp &\sim \sum_{\lambda = -1/2}^{1/2} 
\left(\beta_{+\lambda}^i \beta_{-\lambda}^j - 
\beta_{-\lambda}^i \beta_{+\lambda}^j\right) \cr
&= \left(1+\eta_i \eta_j\right) \beta_{+\lambda}^i 
\beta_{-\lambda}^j.
\end{align}
Thus $d\Delta\sigma_\perp$ vanishes when $\eta_i \eta_j
= -1$. The most dominant trajectory with natural
parity is the proton one. The next one is the excited nucleon $N^*(1520)$ 
of which the spin-parity quantum numbers are given by $J^P=3/2^-$. As
for the unnatural parity states, the inteference between $\Delta$ and
$\Delta(1600)$ exchanges furnishes the most dominant
contribution. 

Then the spin-dependent and spin-averaged differential cross-sections
read 
\begin{align} \label{}
d\Delta\sigma_\perp &= \frac 1{s}
\sum_{i,j} \sum_\lambda 
\beta_{+\lambda}^i \beta_{-\lambda}^j
2\mathrm{Im}\,\mathcal{P}_i \mathcal{P}_j^* \cr
&\times G^{ij}_\mathds{P}(t) \gamma^{pp}_\mathds{P}(0)
\left(\frac{M_X^2}{s_0}\right)^{\alpha_\mathds{P}(0)}, \cr
d\sigma &= \frac 1{s} \sum_{i,j} \sum_\lambda
\beta_{+\lambda}^i \beta_{+\lambda}^j 2\mathrm{Re}\,
\mathcal{P}_i \mathcal{P}_j^* \cr
&\times G^{ij}_\mathds{P}(t) \gamma^{pp}_\mathds{P}(0)
\left(\frac{M_X^2}{s_0}\right)^{\alpha_\mathds{P}(0)},
\end{align}
where the triple-Regge coupling $G^{ij}_\mathds{P}(t)$ is often
parametrized as $G(t)=G(0)e^{bt}$, because it can not be
theoretically determined. In the present work, we parameterize 
the form of the triple-Regge couplings so that we can describe the
RHICf data:
$G^{ii}_\mathds{P}(t) = G^{ii}_\mathds{P}(0) 
e^{-B^{ii}_\mathds{P}|t|}$, $G^{ij}_\mathds{P}(t) = G^
{ij}_\mathds{P}(0) \sqrt{|t|}e^{-B^{ij}_
  \mathds{P}|t|}/m_\pi$.
We define the following parameters 
\begin{align} \label{}
g^{ij}_\mathds{P} \equiv 
G^{ij}_\mathds{P}(0)/G^{NN}_\mathds{P}(0), \;\;\;
b^{ij}_\mathds{P} \equiv B^{ij}_\mathds{P} - B^{NN}_
  \mathds{P}
  \label{eq:12}
\end{align}
and fit them to the RHICf data. In Table~\ref{tab:1}, we list the
numerical values of $g_{\mathds{P}}^{ij}$ and $b_{\mathds{P}}^{ij}$.
\setlength{\tabcolsep}{5pt}
\renewcommand{\arraystretch}{1.5}
\begin{table}[htp]
\caption{Numerical values of the parameters $g_\mathds{P}^{ij}$ and
  $b_\mathds{P}^{ij}$.  The first column lists the values of
  $g_\mathds{P}^{ij}$ with $i$ and $j$ given whereas the second column
  shows the values of $b_\mathds{P}^{ij}$. }
\label{tab:1}
\begin{tabular}{c|cc}
\hline 
\hline 
    & $g^{ij}_\mathds{P}$ & $b^{ij}_\mathds{P}~[
    \mathrm{GeV}^{-2}]$ \\
\hline
$NN^*$ & $0.028$ & $0.2$ \\
$\Delta\Delta^*$ & $-0.018$ & 0 \\
$N^*N^*$ & $0.10$ & 0 \\
$\Delta\Delta$ & $0.022$ & 0 \\
$\Delta^*\Delta^*$ & $0.079$ & 0 \\ 
\hline
\hline 
\end{tabular}%
\end{table}
Note that $b_\mathds{P}^{ij}$ is the subtraction given in
Eq.~\eqref{eq:12}. Except for the $\mathds{P}NN^*$ vertex, all the
values of $b_\mathds{P}^{ij}$ are set to be zero. 

%parameter table
\vspace{0.5em}
\emph{Results and Discussion.}--
\begin{figure}[htp]
\centering
\includegraphics[scale=.6]{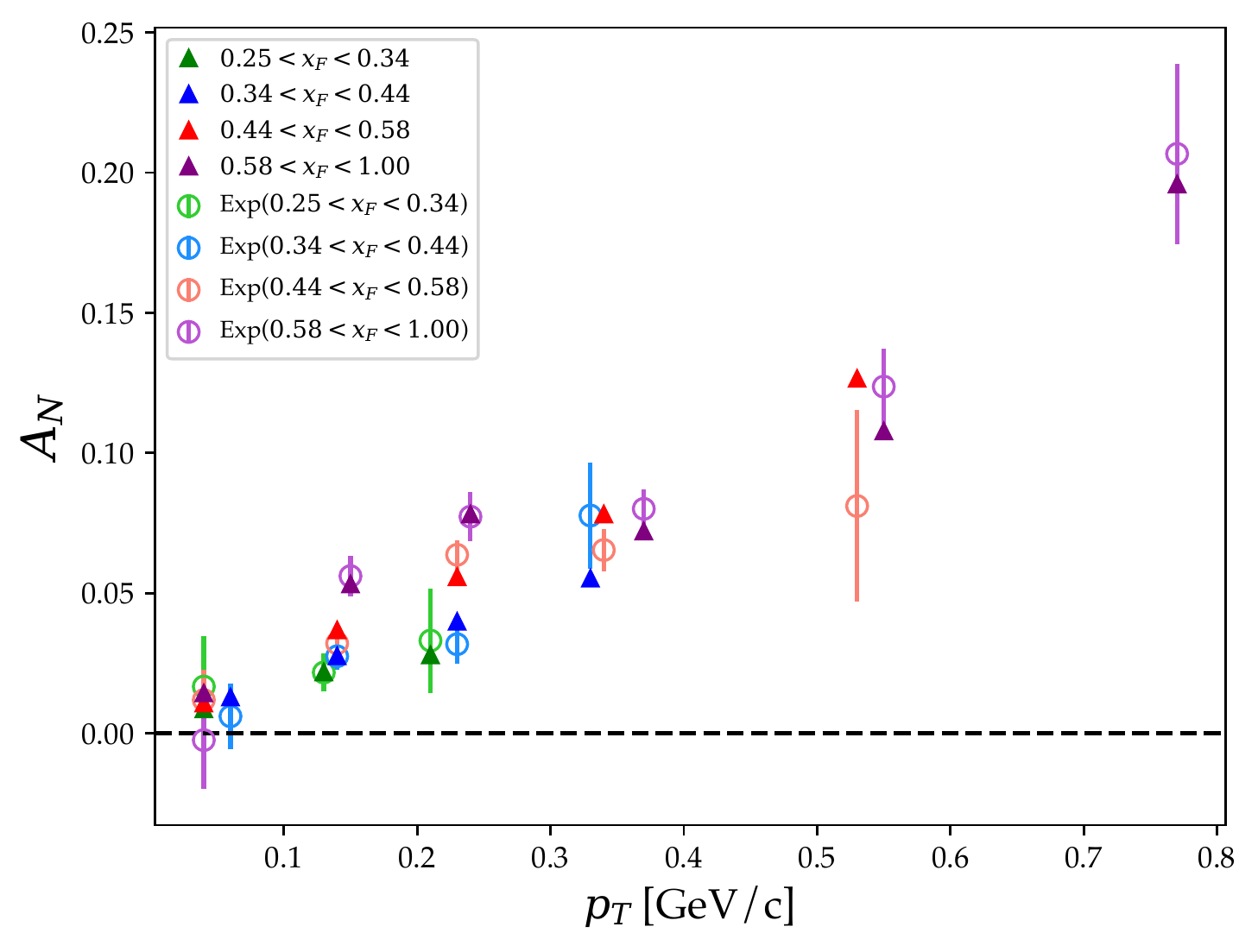}
\caption{Numerical results for the TSSA as a function of $p_T$ with several
ranges of $x_F$ given. The present results are depicted by the
triangles. The open circles with error bars illustrate the RHICf
data~\cite{RHIC-f:2020koe}.}  
\label{fig:2}
\end{figure}
The RHICf Collaboration has first measured $A_N$ for
$p+p^\uparrow\to \pi+X$ as a function of $p_T$ with several different
ranges of $x_F$ given. Since the data on $A_N$ in the negative $x_F$
region are almost equal to zero, we concentrate on $A_N$ with positive
values of $x_F$. In Fig.~\ref{fig:2} we show the numerical results for
$A_N$ given as a function of the transverse momenta
$p_T$ with four different ranges of $x_F$, compared with the RHICf
data~\cite{RHIC-f:2020koe}. The present results are in quantitative
agreement with the data. The value of $A_N$ starts to increase as
$p_T$ increases till $p_T$ reaches $0.2\sim 0.3\,\mathrm{GeV}/c$.
Then, it seems saturated for a while and then enlarges again as $p_T$
further increases. Note that in general the experimental uncertainties
become larger as $p_T$ increases. 

\begin{figure}[htp]
\centering
\includegraphics[scale=.6]{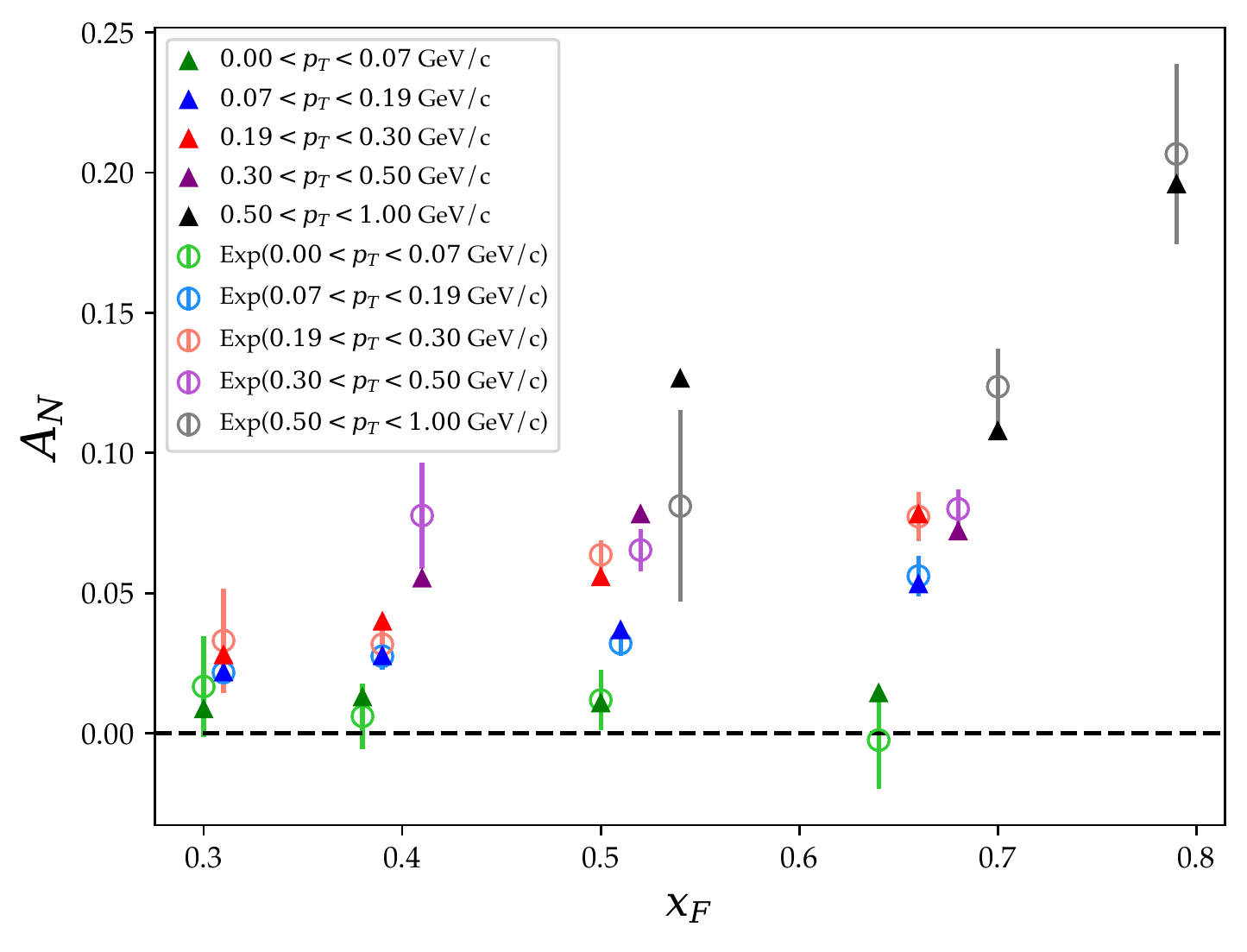}
\caption{The TSSA as a function of $x_F$ with several
ranges of $p_T$ given. Notations are the same as in FIg.~\ref{fig:2}.}
\label{fig:3}
\end{figure}
Figure~\ref{fig:3} displays the numerical results for $A_N$ as a
function of $x_F$ with five different ranges of $p_T$
given~\cite{RHIC-f:2020koe}. The current results have an outstanding
fit with the RHICf data, in particular, as $p_T$ becomes smaller. Note
that when $p_T$ approaches zero, $A_N$ is suppressed.

\begin{figure}[htp]
\centering
\includegraphics[scale=.2]{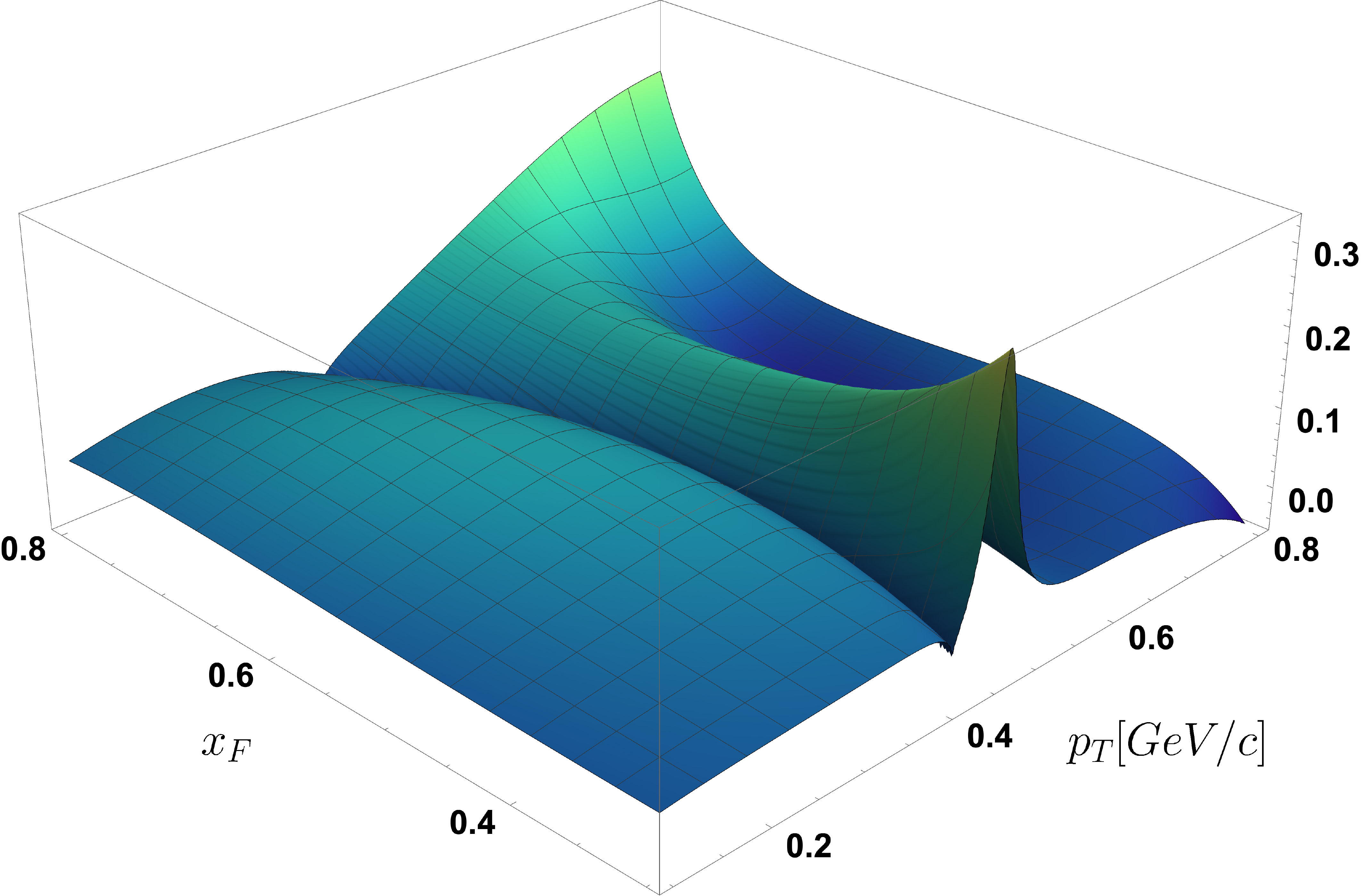}
\caption{The 3d plot of $A_N$ as a function of $p_T$ and
$x_F$.}
\label{fig:4}
\end{figure}

To scrutinize the current results, we plot $A_N$ as a
function of $p_T$ and $x_F$ in Fig~\ref{fig:4}.
It is notable to see the peak in the mid-$p_T$ range ($\sim
0.5\,\mathrm{GeV}/c$), in particular, when $x_F$ is small.
We can understand this feature of $A_N$ by examining the
characteristics of the signature factor. At certain values of $p_T$
and $x_F$, $A_N$ becomes very sensitive to signature factor of the
proton. The peak structure of $A_N$ occurs because of this sensitivity.  
On the other hand, when $x_F$ is large, the diagonal terms such as
$NN$, $N^* N^*$, $\Delta\Delta$, and $\Delta^*\Delta^*$ diagrams come
into play, the peak structure gets smeared. 
As $x_F$ becomes very small ($x_F<0.3$), all the signature factors bring about a
rapid oscillation of $A_N$. It indicates that the current scheme of
the triple-Regge exchange breaks down when $x_F$ is very small. 

\emph{Conclusions.}-- In this work, we aimed at investigating the
transverse single spin asymmetries for the neutral pion production
from inclusive polarized proton and proton collision, emphasizing the
triple-Regge exchange that consists of two baryons and a pomeron.
The numerical results of the current work are in quantitative
agreement with the RHICf data. We discussed the feature of the
transverse single spin asymmetries with $p_T$ and $ x_F$ varied. 
When the pseudorapidity is large and $x_F$ is not very small, we can
interpret the neutral pion production from inclusive polarized proton
and proton collision as a diffractive one. 

\vspace{10pt} 

\begin{acknowledgments}
The authors are grateful to M. H. Kim and B. S. Hong at the 
Center for Extreme Nuclear Matter, Korea University.  
The present work was supported by Basic Science Research Program
through the National Research Foundation of Korea funded by the
Ministry of Education, Science and Technology
(Grant-No. 2021R1A2C209336 and 2018R1A5A1025563). 
\end{acknowledgments}

\end{document}